\documentclass[10pt,twocolumn,aps,pra,amsmath,amssymb,showpacs]{revtex4-1}
\usepackage{bm}
\usepackage{mathrsfs}
\usepackage{graphicx}
\usepackage[usenames,dvipsnames,svgnames,table]{xcolor}
\usepackage{framed}
\usepackage{amsthm}
\usepackage[utf8]{inputenc}

\colorlet{shadecolor}{blue!15}
\usepackage[unicode=true,
            pdfusetitle, 
            bookmarks=true,
            bookmarksnumbered=false,
            bookmarksopen=false,
            breaklinks=true,
            pdfborder={0 0 0},
            backref=false,
            colorlinks=true]{hyperref}
\hypersetup{linkcolor=NavyBlue,urlcolor=NavyBlue,citecolor=NavyBlue}

\usepackage{amsthm}
\newcommand{\propnumber}{} 
\newtheorem*{prop}{Proposition \propnumber}

  \newtheorem{propo}{Proposition} 

 \usepackage{tikz-cd}

\newcommand{\cH}{\mathcal{H}}

\begin{document}

\title {Initial correlations in open quantum
     systems  are always detectable}

\author {Iman Sargolzahi}
\email{sargolzahi@neyshabur.ac.ir; sargolzahi@gmail.com}
\affiliation {Department of Physics, University of Neyshabur, Neyshabur, Iran}


\begin{abstract}
Consider  an open quantum system  which interacts with its environment. Assuming that the experimenter has access only to the system,  an interesting question is  whether it is  possible to detect  initial correlations between the   system and the environment  by  performing measurements only on the system. Various methods have been proposed to detect correlations by local measurements on the system. After reviewing these methods, we will show that initial correlations between the   system and the environment are always detectable. In particular, we will show that one can always find a unitary evolution, for  the whole system-environment, such that the trace distance method, proposed to witness correlations locally,    succeeds. We also find the condition for  existence of the  optimal unitary evolution, for which the entire  correlation is locally detectable. 
Next, we address  the case where the system and the environment interact through a  time-independent  Hamiltonian. For this case, we will see that if the initial correlation can be detected locally at some time $t$, then it can be detected  for almost all the other times too.
On the other hand,  we see that one can find cases for which initial correlations between the system and the environment   always  remain undetectable even though the unitary evolution, generated by the Hamiltonian, is not factorized. 
\end{abstract}

\maketitle

\section{Introduction}


In general, a quantum system $S$ is not closed and  interacts with its environment $E$. We can consider the whole system-environment as a closed quantum system which evolves  unitarily  \cite{1}:
\begin{equation}
\label{eq:1}
 \rho_{SE}^{\prime}= \mathrm{Ad}_U (\rho_{SE}) \equiv U \rho_{SE} U^{\dagger},
 \end{equation} 
 where $U$ is a unitary operator, on $\cH_S\otimes\cH_E$. $\cH_S$ and $\cH_E$ are the Hilbert spaces of the system and the environment, respectively.  In addition, $\rho_{SE}$ and $\rho_{SE}^{\prime}$ are the  initial and the  final states (density operators) of the system-environment, respectively.

So, the reduced dynamics of the system is given by  
\begin{equation}
\label{eq:2}
\rho_{S}^{\prime}=\mathrm{Tr}_{E}(\rho_{SE}^{\prime})=\mathrm{Tr}_{E} \circ \mathrm{Ad}_U (\rho_{SE}).
\end{equation} 
Usually, one assumes that the initial state $\rho_{SE}$ is factorized: $\rho_{SE}=\rho_{S}\otimes \tilde{\omega}_E$, where $\rho_{S}$ is an arbitrary state of the system, but $\tilde{\omega}_E$ is a fixed state of the environment. Therefore, the reduced dynamics of the system is given by a completely positive map:
\begin{equation}
\label{eq:3}
\rho_{S}^{\prime}=\mathcal{E}_S(\rho_S)=\sum_i E_i \rho_{S} E_i^{\dagger}, \qquad   \sum_i E_i^{\dagger} E_i=I_S,
\end{equation}
where $E_i$ are linear operators and $I_S$ is the identity operator on $\cH_S$ \cite{1, 2}.

When the coupling between
the system and environment is weak, the whole state of the system-environment can  remain (approximately) factorized  during the time evolution, and so the reduced dynamics of the system is given by a Markovian master equation \cite{2}. 

Obviously, the above assumption is not valid in general, and so the reduced dynamics may be non-Markovian. To detect and quantify the non-Markovianity some methods have been proposed \cite{3, 4, 5, 6}.
One of them, introduced in Ref. \cite{4}, is based on the trace distance, which is a measure of distinguishability  between two states.
 The trace distance between two states $\rho$ and $\sigma$ is defined as
 \begin{equation}
\label{eq:4}
 D(\rho , \sigma)= \frac{1}{2}\mathrm{Tr}(\vert\rho - \sigma\vert),
\end{equation}
   where $\vert A \vert = \sqrt{A^{\dagger}A}$ \cite{1}.
If, for some time $t$, we have 
 \begin{equation}
\label{eq:5}
\frac{d}{dt} D(\rho_S(t) , \sigma_S(t))> 0,
\end{equation}
for two (reduced) states of the system $\rho_{S}(t)=\mathrm{Tr}_{E}(\rho_{SE}(t))$ and  $\sigma_{S}(t)=\mathrm{Tr}_{E}(\sigma_{SE}(t))$, it means that the distinguishability
between $\rho_{S}(t)$ and  $\sigma_{S}(t)$ is increasing at this moment. This has been interpreted as the consequence of the flow of information from the environment to the system, and thus as the signature of  non-Markovianity \cite{4}.

As the assumption that the whole state of the system-environment remains factorized during the evolution is not valid in general, the assumption that the initial state of the system-environment is factorized  may be violated too. This may be due to the fact that the experimenter could not isolate the system from the environment before beginning the experiment. Therefore, some works have been focused on describing the reduced dynamics of the system when the initial states of the system-environment are not necessarily factorized (see, e.g., Refs.  \cite{7, 8, 9, 10, 11, 12}).

An interesting question is    whether it is  possible to detect the  initial correlation between the   system and the environment  by only tracking the (reduced) dynamics of the system.
In general, the experimenter has access only to the system, and not to the environment. Even if the experimenter  has access  to  both the system and  the environment, obviously, performing  state tomography  on the system is much simpler than doing so on the whole system-environment. Therefore, it would be desirable to get information about the whole system-environment by performing measurements only on the system.

Interestingly, the evolution of the trace distance can be used to answer the above question too \cite{13, 3}.
For two initial states of the system-environment  $\rho_{SE}=\rho_{S}\otimes \tilde{\omega}_E$ and $\sigma_{SE}=\sigma_{S}\otimes \tilde{\omega}_E$, using Eq. \eqref{eq:2}, the reduced dynamics of the system is given by the completely positive map in Eq  \eqref{eq:3}. Now, since the trace distance is contractive under completely positive maps \cite{1}, we have
 \begin{equation}
\label{eq:6}
 D(\mathcal{E}_S(\rho_S) , \mathcal{E}_S(\sigma_S))\leq D(\rho_S , \sigma_S).
\end{equation}
So, if one observes an increase of the trace distance, above the initial value $D(\rho_S , \sigma_S)$, this may imply that, at least, one of the initial states $\rho_{SE}$ or $\sigma_{SE}$ is not factorized.
In the next section we  review this method in more detail, and also other proposed methods to detect  initial correlations by performing measurements only on the system.

In this paper we  show that if there exists  initial correlation between the system and the environment, it is always detectable by  tracking only the reduced dynamics of the system. In other words, there exists a unitary time evolution $U$, for the whole system-environment, such that the reduced dynamics of the system for the  correlated  initial state of the system-environment differs  from that of the  factorized one. So the initial correlation can be detected using the trace distance method. This result is given in Sec. \ref{sec: C}.

Next, in Sec. \ref{sec: D} we give some more results on the trace distance methods of detecting initial correlation, reviewed  in Sec. \ref{sec:B}. 
 The possibility of finding the optimal $U$ and also the case that the system-environment evolution is governed by a time-independent Hamiltonian are studied in Secs. \ref{sec: E} and \ref{sec: F}, respectively. 
 Finally, we end this paper in  Sec. \ref{sec: G}, with a summary  of our results.

\section{Various methods of detecting initial correlations}  \label{sec:B}


As stated in the Introduction, one proposed way to detect  initial correlations between the system and the environment is to track the trace distance of the reduced states. Consider two different initial states of the system-environment $\rho_{SE}$ and $\sigma_{SE}$. The following inequality has been proved in Ref. \cite{13}:
\begin{equation}
\label{eq:7}
\begin{aligned}
 D(\rho_{S}^\prime , \sigma_{S}^\prime) - D(\rho_{S} , \sigma_{S}) \leq \ D(\rho_{SE}, \rho_{S}\otimes \rho_{E})  \qquad \\
 + \ D(\sigma_{SE}, \sigma_{S}\otimes \sigma_{E}) \quad \\ + \ D(\rho_{E} , \sigma_{E}), \qquad\quad \
\end{aligned}
\end{equation}
where $\rho_{S}=\mathrm{Tr}_{E}(\rho_{SE})$ and $\sigma_{S}=\mathrm{Tr}_{E}(\sigma_{SE})$ are the initial states of the system, and  $\rho_{E}=\mathrm{Tr}_{S}(\rho_{SE})$ and $\sigma_{E}=\mathrm{Tr}_{S}(\sigma_{SE})$ are the initial states of the environment. In addition, $\rho_{S}^\prime=\mathrm{Tr}_{E}(U \rho_{SE} U^\dagger)$ and $\sigma_{S}^\prime=\mathrm{Tr}_{E}(U \sigma_{SE} U^\dagger)$ are the final  states of the system.
Therefore, if the trace distance of the reduced states of the system increases after the evolution, it implies that $\rho_{SE}$ or $\sigma_{SE}$ is correlated, or  $\rho_{E}$ differs from $\sigma_{E}$.  One can also find a generalization of inequality \eqref{eq:7} using another quantifier of distinguishability instead of the trace distance \cite{14}. 

Inequality \eqref{eq:7} can be used to detect  correlation in an unknown initial state $\rho_{SE}$, as  follows \cite{13}: Construct the state $\sigma_{SE}$  by performing a completely positive map $\mathcal{F}_S$    on the system, $\sigma_{SE}= \mathcal{F}_S \otimes  \mathrm{id}_E (\rho_{SE})$,   where $\mathrm{id}_E$ is the identity map on the environment.
Therefore, $\sigma_{E} = \rho_{E}$, and if $\rho_{SE}$  is factorized, so is $\sigma_{SE}$. Consequently, any increase in the trace distance implies that the initial  $\rho_{SE}$ is correlated.

Note that the above method includes no restrictions on the initial state $\rho_{SE}$ , system-environment evolution $U$, and the quantum operation on the system $\mathcal{F}_S$. It only requires that the experimenter can perform quantum operations and state tomography on the system $S$.
Therefore this method is experimentally feasible and has been implemented successfully, detecting the initial correlations \cite{15, 16}.

There are at least two interesting choices for $\mathcal{F}_S$. First, if one chooses $\mathcal{F}_S$ as the measurement in the basis of the eigenstates of initial $\rho_{S}$, 
then it can be shown that any increase in the trace distance is a witness that the initial $\rho_{SE}$ includes quantum correlation \cite{17}, i.e., it includes  quantum discord, introduced in Ref. \cite{18}.
Second, when one chooses $\mathcal{F}_S$ such that
\begin{equation}
\label{eq:8}
\begin{aligned}
\sigma_{SE}= \mathcal{F}_S \otimes  \mathrm{id}_E (\rho_{SE})= \rho_{S}\otimes \rho_{E}.
\end{aligned}
\end{equation} 
We can construct $\mathcal{F}_S$ as $\mathcal{F}_S= \Lambda_S \circ \vert 0_S\rangle\langle 0_S \vert \mathrm{Tr}_{S}$, where $ \vert 0_S\rangle \in \cH_S$ is a fixed state and  $\Lambda_S$ is a completely positive map that maps $\vert 0_S\rangle\langle 0_S \vert$ to $\rho_S$. Given two arbitrary density operators, one can always find a completely positive map which maps one to the other \cite{19}.

For $\sigma_{SE}$ in Eq. \eqref{eq:8},   inequality \eqref{eq:7} is simplified as
\begin{equation}
\label{eq:9}
 D(\rho_{S}^\prime , \sigma_{S}^\prime)  \leq \ D(\rho_{SE}, \rho_{S}\otimes \rho_{E}).
\end{equation}
So, if $\rho_{S}^\prime \neq \sigma_{S}^\prime$, we conclude that the initial $\rho_{SE}$ is not factorized. In the next section, we will show that, for any correlated $\rho_{SE}$, one can find unitary evolution $U$ such that we have $\rho_{S}^\prime \neq \sigma_{S}^\prime$.

Interestingly, the trace distance approach can be used to detect correlations within the environment too \cite{23}.
Consider the case that the environment $E$ is bipartite: $\cH_E=\cH_B \otimes \cH_C$. For the two initial states of the system-environment  $\rho_{SE}=\rho_{S} \otimes \rho_{E}=\rho_{S} \otimes \rho_{BC}$ and  $\sigma_{SE}=\sigma_{S} \otimes \rho_{B} \otimes \rho_{C}$, where  $\rho_{B}=\mathrm{Tr}_{C}(\rho_{BC})$ and $\rho_{C}=\mathrm{Tr}_{B}(\rho_{BC})$, the following inequality has been proven in Ref. \cite{23}:
\begin{equation}
\label{eq:10}
\begin{aligned}
 D(\rho_{S}^\prime , \sigma_{S}^\prime) - D(\rho_{S} , \sigma_{S}) \leq \ D(\rho_{BC}, \rho_{B}\otimes \rho_{C}). 
\end{aligned}
\end{equation}
Therefore,  an increase in the trace distance implies that in the initial state $\rho_{SE}$  the environment is not factorized.

In many cases  it is known how
 the system and the environment interact. This can help with detecting initial correlations, as  is shown in Refs. \cite{20, 21}, and is implemented experimentally in Ref. \cite{22}.
For example, in Ref. \cite{20} the authors have considered the case that the system-environment Hamiltonian is known. So the reduced dynamics of the system can be calculated simply for all factorized initial states $\bar{\rho}_{SE} =\rho_S \otimes \bar{\rho}_E$, where $\rho_S$ is a given initial state of the system but $\bar{\rho}_E$ is an arbitrary state of the environment. Then, these calculated final states of the system can be compared with the real one  $\rho_{S}^\prime$, given from the state tomography. If they do not match, we conclude that the initial state of the system-environment $\rho_{SE}$ is not factorized as $\bar{\rho}_{SE}$.

Master equations can give us some information about the system-environment correlations too. In Ref. \cite{24} it has been shown that, knowing $\rho_{S}(t)$ (from solving the master equation), the system-environment interaction, and also the state of the environment at least at the initial moment $t=0$, gives us the approximate system-environment correlation at the other  times $t$.

Finally, we mention that the initial correlation can  remarkably  affect the quantum process tomography \cite{26}. 
Consider the following method for the process tomography \cite{27}: The experimenter,  who has access only to the system, starts with an unknown (maybe correlated) initial state of the system-environment $\rho_{SE}^{(1)}$. Implementing a  process  $\mathcal{E}_S$ on the system, given by an unknown unitary operator  $U$ on the whole system-environment, the final state of the system becomes  ${\rho_{S}^\prime}^{(1)}= \mathcal{E}_S (\rho_{S}^{(1)}) =\mathrm{Tr}_{E}(U\rho_{SE}^{(1)}U^\dagger)$, where $\rho_{S}^{(1)}=\mathrm{Tr}_{E}(\rho_{SE}^{(1)})$ is the initial state of the system. The experimenter  characterizes both $\rho_{S}^{(1)}$ and ${\rho_{S}^\prime}^{(1)}$   by performing state tomography \cite{1}.

Assuming that the system is $d_S$-dimensional, in addition to  $\rho_{S}^{(1)}$, one can find $(d_S^2-1)$ other linearly independent states $\rho_{S}^{(i)}$, $i \neq 1$, such that the set $\lbrace\rho_{S}^{(1)}, \rho_{S}^{(2)}, \dots, \rho_{S}^{(d_S^2)} \rbrace$ constructs  a basis for the space of the  linear operators on $\cH_S$. So, any arbitrary state  $\rho_{S}$ on $\cH_S$ can be expanded as
\begin{equation}
\label{eq:11}
\begin{aligned}
\rho_{S}=\sum_{i=1}^{d_S^2} a_i \rho_{S}^{(i)},
\end{aligned}
\end{equation}
where $a_i$ are real coefficients.

The  experimenter can construct each $\rho_{S}^{(i)}$ by performing a suitable completely positive map $\mathcal{F}_S^{(i)}$ on $\rho_{S}^{(1)}$ \cite{19}, so the whole state of the system-environment converts to $\rho_{SE}^{(i)}=\mathcal{F}_S^{(i)} \otimes  \mathrm{id}_E  (\rho_{SE}^{(1)})$.
 Then, implementing the  mentioned unknown process $\mathcal{E}_S$ on this new initial state, he/she can find the corresponding final state ${\rho_{S}^\prime}^{(i)}= \mathcal{E}_S(\rho_{S}^{(i)})=\mathrm{Tr}_{E}(U\rho_{SE}^{(i)}U^\dagger)$.

Until now, we know how $\mathcal{E}_S$ acts on $\rho_{S}^{(i)}$, $i=1, \dots, d_S^2$.
Assuming that $\mathcal{E}_S$ is linear, and using Eq. \eqref{eq:11}, we can obtain the final state of the system for the arbitrary initial state $\rho_{S}$:
\begin{equation}
\label{eq:12}
\begin{aligned}
\rho_{S}^\prime= \mathcal{E}_S(\rho_{S})=\sum_{i=1}^{d_S^2} a_i \mathcal{E}_S(\rho_{S}^{(i)})
=\sum_{i=1}^{d_S^2} a_i {\rho_{S}^\prime}^{(i)}.
\end{aligned}
\end{equation}
In other words, Eq. \eqref{eq:12} determines how the quantum process  $\mathcal{E}_S$ acts on an  arbitrary initial state $\rho_{S}$.

If the initial  $\rho_{SE}^{(1)}$ is factorized as $\rho_{SE}^{(1)}=\rho_{S}^{(1)}\otimes \tilde{\omega}_E$, and so the other  $\rho_{SE}^{(i)}$ are also factorized as $\rho_{SE}^{(i)}=\rho_{S}^{(i)}\otimes \tilde{\omega}_E$, then the above method for the  process tomography works well. In fact, then  $\mathcal{E}_S$ is not only linear but is also completely positive. But, if  $\rho_{SE}^{(1)}$ is not factorized, Eq. \eqref{eq:12} may fail in predicting the final $\rho_{S}^\prime$ correctly \cite{26}. We will come back to this point in the next section too.

\section{Initial correlation is detectable}  \label{sec: C}

Consider two initial states of the system-environment $\rho_{SE}$ and $\sigma_{SE}$, such that $\rho_{S}=\mathrm{Tr}_{E}(\rho_{SE})=\mathrm{Tr}_{E}(\sigma_{SE})=\sigma_S$. In other words, though $\rho_{SE}$ differs from 
$\sigma_{SE}$,  the initial state of the system  is the same for both $\rho_{SE}$ and $\sigma_{SE}$.  
So,
 \begin{equation}
\label{eq:13}
\begin{aligned}
\sigma_{SE}= \rho_{SE}+ R,
\end{aligned}
\end{equation}
where $R$ is a Hermitian operator, on $\cH_S\otimes\cH_E$, such that $\mathrm{Tr}_{E}(R)=0$. 
We now ask  whether the final state of the system is also the same for both initial states  $\rho_{SE}$ and $\sigma_{SE}$. In  the following Proposition, we  show that one can always find a unitary evolution $U$, for the whole system-environment, such that 
 $\sigma_{S}^{\prime}=\mathrm{Tr}_{E} (U \sigma_{SE} U^{\dagger})$  differs from  $\rho_{S}^{\prime}$ in Eq. \eqref{eq:2}.
\begin{propo}
\label{pro1}
Consider the case that both the system and the environment are finite dimensional, with  dimensions  $d_S \geq 2$ and $d_E \geq 2$, respectively.
  For the two initial states of the system-environment $\rho_{SE}$ and $\sigma_{SE}$ in Eq. \eqref{eq:13},
 one can always find a unitary operator $U$
   such that  $\mathrm{Tr}_{E}(URU^\dagger) \neq 0$. 
 \end{propo}
\textit{Proof.} The  Hermitian traceless operator $R$ can be expanded as
\begin{equation}
\label{eq:14}
\begin{aligned}
R=\sum_{i=1}^n \mu_i^{(+)}  \vert \mu_i^{(+)}\rangle\langle \mu_i^{(+)}\vert +  \sum_{i=n+1}^N  \mu_i^{(-)}  \vert \mu_i^{(-)}\rangle\langle \mu_i^{(-)}\vert,
\end{aligned}
\end{equation}
where $\mu_i^{(+)}$ and $\mu_i^{(-)}$ are positive and negative eigenvalues of $R$, respectively. The states $\vert \mu_i^{(+)}\rangle$ and $\vert \mu_i^{(-)}\rangle$ are the corresponding eigenstates. In addition, $N=d_S d_E$.
Obviously, $n=m d_E + r$, with the integers $0\leq m<d_S$ and $0\leq r<d_E$. We add the zero eigenvalues of $R$ to the sets  $\lbrace\mu_i^{(+)}\rbrace$ or $\lbrace\mu_i^{(-)}\rbrace$ appropriately, such that  $r$ is as small as possible.

If $m \geq 1$, we consider a unitary operator $U$ such that, for $m d_E$ of $\vert \mu_i^{(+)}\rangle$, we have
\begin{equation}
\label{eq:15}
\begin{aligned}
U \vert \mu_i^{(+)}\rangle = \vert j_S \rangle \vert l_E\rangle,
\end{aligned}
\end{equation}
where the states $\vert j_S \rangle$,  $1 \leq j \leq m$, are some members of an orthonormal basis of $\cH_S$, and 
the states $\vert l_E \rangle$,  $1 \leq l \leq d_E$, construct an orthonormal basis of $\cH_E$.
In addition, $U$ maps the $r$ remaining $\vert \mu_i^{(+)}\rangle$ and all the  $\vert \mu_i^{(-)}\rangle$
to $\vert j_S \rangle \vert l_E\rangle$, with $m+1 \leq j \leq d_S$ and $1 \leq l \leq d_E$.

Therefore,
\begin{equation}
\label{eq:15a}
\begin{aligned}
R^\prime=URU^\dagger =R^{\prime (p)} + R^{\prime (h)},
\end{aligned}
\end{equation}
where $R^{\prime (p)}$ is a positive operator on the subspace spanned by the states $\vert j_S \rangle \vert l_E\rangle$, with $1 \leq j \leq m$ and  $1 \leq l \leq d_E$. In addition, $R^{\prime (h)}$ is a Hermitian operator on the subspace spanned by  $\vert j_S \rangle \vert l_E\rangle$,  with $m+1 \leq j \leq d_S$ and $1 \leq l \leq d_E$. Tracing over the environment, we have
\begin{equation}
\label{eq:15b}
\begin{aligned}
R^\prime_S=\mathrm{Tr}_{E} (R^\prime) =R^{\prime (p)}_S + R^{\prime (h)}_S,
\end{aligned}
\end{equation}
where $R^{\prime (p)}_S=\mathrm{Tr}_{E} (R^{\prime (p)})$ is a positive operator on the subspace spanned by  $\vert j_S \rangle$,  $1 \leq j \leq m$, and $R^{\prime (h)}_S=\mathrm{Tr}_{E} (R^{\prime (h)})$ is a Hermitian operator on the subspace spanned by  $\vert j_S \rangle$,  $m+1 \leq j \leq d_S$.

 Since the positive operator  $R^{\prime (p)}_S$ is nonzero and its support does not overlap with the support of $R^{\prime (h)}_S$, the Hermitian  operator $R_S^\prime$ in Eq. \eqref{eq:15b} is also nonzero. (In fact, since $R_S^\prime$ is traceless, the Hermitian  operator $R^{\prime (h)}_S$ is nonzero too.)
 
%

If  $m=0$, we can follow a similar line of reasoning for  $\vert \mu_i^{(-)}\rangle$ instead of  $\vert \mu_i^{(+)}\rangle$ and conclude that the operator  $R_S^\prime$ is nonzero. $\qquad\qquad\qquad\qquad\qquad\qquad\qquad\qquad\quad\blacksquare$

Let us illustrate the above proof for the simple case where $d_S=2$, $m=1$ and $r=0$. So, $R$ in Eq. \eqref{eq:14} is
\begin{equation}
\label{eq:14a}
\begin{aligned}
R=\sum_{i=1}^{d_E} \mu_i^{(+)}  \vert \mu_i^{(+)}\rangle\langle \mu_i^{(+)}\vert +  \sum_{i=d_E+1}^{2d_E} \mu_i^{(-)}  \vert \mu_i^{(-)}\rangle\langle \mu_i^{(-)}\vert.
\end{aligned}
\end{equation}
We choose $U$ such that it maps all the kets $ \vert \mu_i^{(+)}\rangle$ to $\vert 1_S \rangle \vert l_E\rangle$, and all the kets $ \vert \mu_i^{(-)}\rangle$ to $\vert 2_S \rangle \vert l_E\rangle$  ($1 \leq l \leq d_E$).
Therefore, in Eq. \eqref{eq:15b}, $R^{\prime (p)}_S=(\sum \mu_i^{(+)}) \vert 1_S\rangle\langle 1_S\vert$  and $R^{\prime (h)}_S=(\sum \mu_i^{(-)}) \vert 2_S\rangle\langle 2_S\vert$.
Obviously, all the operators $R^{\prime (p)}_S$, $R^{\prime (h)}_S$ and $R^{\prime }_S$ are nonzero, as expected.

In addition, note that one can find infinitely many unitary operators $U$ which satisfy Eq. \eqref{eq:15}. In other words, there are infinitely many unitary evolutions $U$, for the whole system-environment, such that the reduced dynamics of the system,  for  the initial states  $\rho_{SE}$ and $\sigma_{SE}$ in Eq. \eqref{eq:13}, is not the same.

For an arbitrary initial state $\rho_{SE}$ we can construct $\sigma_{SE}$ as Eq. \eqref{eq:8}. When $\rho_{SE}$ is correlated, i.e., $\rho_{SE} \neq\rho_{S}\otimes \rho_{E}$, then $R$ in Eq. \eqref{eq:13} is nonzero. Using Proposition \ref{pro1}, we conclude that one can find unitary evolution $U$ such that $\rho_{S}^\prime \neq \sigma_{S}^\prime$, and so inequality \eqref{eq:9} can detect  correlation in the initial state $\rho_{SE}$. In other words,  one can  always find an appropriate $U$ such that the method introduced in Ref. \cite{13} to detect the initial correlation is successful.

We can also apply  Proposition \ref{pro1} to realize why  the  process tomography method, given in the previous section, may fail.
First note that for the $\rho_{S}=\mathrm{Tr}_{E}(\rho_{SE})$ expanded in Eq. \eqref{eq:11}, we have
\begin{equation}
\label{eq:16}
\begin{aligned}
\rho_{SE}=\sum_{i=1}^{d_S^2} a_i \rho_{SE}^{(i)} + Y,
\end{aligned}
\end{equation}
where $Y$ is a Hermitian operator, on $\cH_S\otimes\cH_E$, such that $\mathrm{Tr}_{E}(Y)=0$.
Now, it can be shown simply that for the system-environment unitary evolution $U$, the reduced dynamics 
$\mathcal{E}_S$ is linear if and only if $\mathrm{Tr}_{E}(UYU^\dagger) = 0$ \cite{12}.
From Eq. \eqref{eq:12} it is obvious  that assuming  $\mathcal{E}_S$ is linear is the main assumption of the mentioned process tomography method. If this assumption is failed, the mentioned method is failed too.

Using a similar line of reasoning given in  Proposition \ref{pro1}, we can show that if $Y$ is nonzero, one can always find a unitary operator $U$ such that $\mathrm{Tr}_{E}(UYU^\dagger)$ is also nonzero.
So, the reduced dynamics  $\mathcal{E}_S$ is not linear, and the mentioned process tomography method will fail.

Now we can show simply that when the initial $\rho_{SE}^{(1)}$ is not factorized, then the mentioned  method may fail. Remember,  we have assumed that the experimenter can  manipulate only
 the system, and not the environment. 
So, using the local operation $\mathcal{F}_S$ in Eq. \eqref{eq:8}, he/she can construct the state $\sigma_{SE}=\rho_{S}^{(1)}\otimes\rho_{E}^{(1)}$, where $\rho_{E}^{(1)}= \mathrm{Tr}_{S} (\rho_{SE}^{(1)})$.
Using Eq. \eqref{eq:11}, we can expand $\sigma_{S}= \mathrm{Tr}_{E} (\sigma_{SE})$ as $\sigma_{S}=\rho_{S}^{(1)}$. So, using Eq. \eqref{eq:16}, $\sigma_{SE}=\rho_{SE}^{(1)} + Y$, where $Y \neq 0$.
Therefore, for each unitary evolution $U$ which leads to a  nonzero $\mathrm{Tr}_{E}(UYU^\dagger)$, the mentioned process tomography method will fail.

In summary, construct the set of   initial states of the  system-environment  as
\begin{equation}
\label{eq:17}
\begin{aligned}
\mathcal{S}= \lbrace \rho_{SE}=\mathcal{F}_S \otimes  \mathrm{id}_E (\rho_{SE}^{(1)})\rbrace ,
\end{aligned}
\end{equation}
where $\mathcal{F}_S$ are  arbitrary completely positive maps on the system. Then, expand each $\rho_{SE} \in \mathcal{S}$ as Eq.  \eqref{eq:16}. If, for a given $U$, $\mathrm{Tr}_{E}(UYU^\dagger)=0$ for all  $\rho_{SE} \in \mathcal{S}$, then the reduced dynamics  $\mathcal{E}_S$ is linear \cite{12} and the mentioned process tomography method will work; otherwise, it may fail.

\section{More on the trace distance methods of detecting initial correlations}  \label{sec: D}

When the initial state $\rho_{SE}$ is correlated, then  $R=\rho_{S}\otimes \rho_{E}- \rho_{SE} \neq 0$.
Now, 
 from the left-hand side of  inequality \eqref{eq:9}, it is clear that     $R_S^\prime=\mathrm{Tr}_{E}(URU^\dagger)\neq 0$ 
is the necessary and sufficient condition that this inequality can detect  correlation.
In the following we want to prove similar results for inequalities \eqref{eq:7} and \eqref{eq:10}.

First, consider inequality \eqref{eq:7}.
If we want to detect correlation in the initial state $\rho_{SE}$, using the method introduced in Ref. \cite{13}, we have  $\sigma_{SE}= \mathcal{F}_S \otimes  \mathrm{id}_E (\rho_{SE})$, performing some quantum operation $ \mathcal{F}_S$ on the system. Therefore  $\rho_{E}=\sigma_{E}$, and so the last term in the right-hand side of inequality \eqref{eq:7} vanishes. 
When the initial $\rho_{SE}$ is correlated, then  $R=\rho_{S}\otimes \rho_{E}- \rho_{SE} \neq 0$. We may also have $\bar{R}=\sigma_{S}\otimes \sigma_{E}- \sigma_{SE} \neq 0$.

Note that since $\rho_{E}=\sigma_{E}$, we have
\begin{equation}
\label{eq:18}
\begin{aligned}
\rho_{SE}- \sigma_{SE}= \bar{R} - R + Q,
\end{aligned}
\end{equation}
where
 $Q=\rho_{S}\otimes \rho_{E} - \sigma_{S}\otimes \rho_{E}$.
In addition, using Eq. \eqref{eq:4}, we have
\begin{equation}
\label{eq:19}
\begin{aligned}
 D(\rho_S , \sigma_S)= \frac{1}{2}\mathrm{Tr}(\vert Q\vert)= \frac{1}{2}\mathrm{Tr}(\vert UQU^\dagger \vert) \quad \\
  \geq \frac{1}{2}\mathrm{Tr}(\vert \mathrm{Tr}_E (UQU^\dagger) \vert). \qquad
\end{aligned}
\end{equation}
Performing the operation $\mathrm{Tr}_{E} \circ \mathrm{Ad}_U $ to  both sides of Eq. \eqref{eq:18}, and then using the triangle inequality for the trace distance \cite{1}, we have
\begin{equation}
\label{eq:20}
\begin{aligned}
 D(\rho_S^\prime , \sigma_S^\prime)\leq   \frac{1}{2}\mathrm{Tr}(\vert \mathrm{Tr}_E (URU^\dagger) \vert) +  \frac{1}{2}\mathrm{Tr}(\vert \mathrm{Tr}_E (U\bar{R}U^\dagger) \vert) \\
 +  \frac{1}{2}\mathrm{Tr}(\vert \mathrm{Tr}_E (UQU^\dagger) \vert) \qquad\qquad\qquad\qquad \\
 = \frac{1}{2}\mathrm{Tr}(\vert R_S^\prime \vert ) +  \frac{1}{2}\mathrm{Tr}(\vert \bar{R}_S^\prime \vert ) \qquad\qquad\qquad\qquad \\ 
  +    \frac{1}{2}\mathrm{Tr}(\vert \mathrm{Tr}_E (UQU^\dagger) \vert).  \qquad\qquad\qquad\qquad
\end{aligned}
\end{equation}
Now, subtracting Eq. \eqref{eq:19} from Eq. \eqref{eq:20}, we conclude that
\begin{equation}
\label{eq:21}
\begin{aligned}
 D(\rho_S^\prime , \sigma_S^\prime) - D(\rho_S , \sigma_S) \leq \frac{1}{2}\mathrm{Tr}(\vert R_S^\prime \vert ) +  \frac{1}{2}\mathrm{Tr}(\vert \bar{R}_S^\prime \vert ).
\end{aligned}
\end{equation}
Therefore, in order that inequality \eqref{eq:7} can detect correlation in the initial state $\rho_{SE}$, at least, one of the Hermitian operators $R_S^\prime$ or $\bar{R}_S^\prime$ must be nonzero.
We will use this result in Sec. \ref{sec: F}.

Next, consider inequality \eqref{eq:10}.
If we choose  $\sigma_{SE}$ as   $\sigma_{SE}=\rho_{S} \otimes \rho_{B} \otimes \rho_{C}$, then inequality \eqref{eq:10} reads 
\begin{equation}
\label{eq:22}
\begin{aligned}
 D(\rho_{S}^\prime , \sigma_{S}^\prime)  \leq \ D(\rho_{BC}, \rho_{B}\otimes \rho_{C}). 
\end{aligned}
\end{equation}
As Eq. \eqref{eq:13}, we can define $R=\sigma_{SE}- \rho_{SE}= \rho_{S} \otimes \rho_{B} \otimes \rho_{C} - \rho_{S} \otimes \rho_{BC}$. Thus  $\mathrm{Tr}_E (R)=0$.
Now, inequality \eqref{eq:22} states that if  $R_S^\prime=\mathrm{Tr}_{E}(URU^\dagger)\neq 0$, we can detect  correlation in the state of environment $\rho_{E}=\rho_{BC}$. From Proposition \ref{pro1}, we know that one can find such unitary evolution $U$, and so the  correlation in the  environment is always detectable.

For the general $\sigma_{SE}$ in inequality \eqref{eq:10}, i.e., when  $\sigma_{SE}=\sigma_{S} \otimes \rho_{B} \otimes \rho_{C}$, we can define the Hermitian operator $Q= \rho_{S} \otimes \rho_{B} \otimes \rho_{C} -\sigma_{S} \otimes \rho_{B} \otimes \rho_{C}$. Thus inequality \eqref{eq:19} is valid for this case too. Also, we have a similar equation as Eq. \eqref{eq:18}, with  $R= \rho_{S} \otimes \rho_{B} \otimes \rho_{C} - \rho_{S} \otimes \rho_{BC}$ and $\bar{R}=0$. Therefore, for inequality \eqref{eq:10}, we have a similar inequality as given in Eq. \eqref{eq:21}:
\begin{equation}
\label{eq:23}
\begin{aligned}
 D(\rho_S^\prime , \sigma_S^\prime) - D(\rho_S , \sigma_S) \leq \frac{1}{2}\mathrm{Tr}(\vert R_S^\prime \vert ).
\end{aligned}
\end{equation}
So, in order that inequality \eqref{eq:10} can detect correlation in the environment, we must have 
$R_S^\prime=\mathrm{Tr}_{E}(URU^\dagger)\neq 0$.

In summary, the nonzero-ness of the Hermitian operator $R_S^\prime$ is the necessary and sufficient condition for applicability of inequality \eqref{eq:10} to detect initial correlation in the environment.

\section{Finding the optimal unitary evolution}  \label{sec: E}

In this section we want to consider inequality \eqref{eq:9} and find the condition for which the equality sign holds in this relation.

First, we expand the Hermitian traceless operator $R=\rho_{S}\otimes \rho_{E}- \rho_{SE}$ as Eq. \eqref{eq:14}. So, using Eq. \eqref{eq:4}, we have
\begin{equation}
\label{eq:24}
 D(\rho_{SE}, \rho_{S}\otimes \rho_{E})=\frac{1}{2}\mathrm{Tr}(\vert R \vert )=\sum_{i=1}^n \mu_i^{(+)}.
\end{equation}
After performing the unitary evolution $U$, we have
\begin{equation}
\label{eq:25}
R^\prime =URU^\dagger = \Gamma - \Delta. 
\end{equation}
The positive operators  $\Gamma$  and $\Delta$ are defined as
\begin{equation}
\label{eq:26}
\begin{aligned}
\Gamma=\sum_{i=1}^n \mu_i^{(+)}  \vert \hat{\mu}_i^{(+)}\rangle\langle \hat{\mu}_i^{(+)}\vert,  \qquad \\ 
 \Delta= -\sum_{i=n+1}^N  \mu_i^{(-)}  \vert \hat{\mu}_i^{(-)}\rangle\langle \hat{\mu}_i^{(-)}\vert,
\end{aligned}
\end{equation}
where $\vert \hat{\mu}_i^{(\pm)}\rangle =U \vert \mu_i^{(\pm)}\rangle$. Therefore, $R^\prime_S=\mathrm{Tr}_E (R^\prime)= \Gamma_S - \Delta_S$, where  $\Gamma_S= \mathrm{Tr}_E (\Gamma)$ and $\Delta_S= \mathrm{Tr}_E (\Delta)$.

Consider the case that $n$, i.e., the number of positive eigenvalues of $R$, is $n=m d_E$, for some integer $0<m<d_S$. So, we can choose $U$ as Eq. \eqref{eq:15}, and conclude that the left hand side of inequality \eqref{eq:9} reads
\begin{equation}
\label{eq:27}
\begin{aligned}
 D(\rho_S^\prime , \sigma_S^\prime) = \frac{1}{2}\mathrm{Tr}(\vert R_S^\prime \vert )=\mathrm{Tr}(  \Gamma_S )=\sum_{i=1}^n \mu_i^{(+)},
\end{aligned}
\end{equation}
since the supports of  $\Gamma_S$ and  $\Delta_S$ are orthogonal. Therefore we achieve the equality sign in Eq. \eqref{eq:9}.

In general, the supports of  $\Gamma_S$ and  $\Delta_S$ are orthogonal if we can decompose the Hilbert space of the system as 
\begin{equation}
\label{eq:28}
\begin{aligned}
\cH_S=\cH_S^{(+)} \oplus \cH_S^{(-)},
\end{aligned}
\end{equation}
such that all $\vert \hat{\mu}_i^{(+)}\rangle \in \cH_S^{(+)} \otimes\cH_E$ and all $\vert \hat{\mu}_i^{(-)}\rangle \in \cH_S^{(-)} \otimes\cH_E$.
 Obviously, this will be the case only when $n=m d_E$. Otherwise, we cannot find any  unitary $U$ which leads to orthogonal  supports for  $\Gamma_S$ and  $\Delta_S$.

Note that, in general, we have 
\begin{equation}
\label{eq:29}
\begin{aligned}
 D(\rho_S^\prime , \sigma_S^\prime) = \frac{1}{2}\mathrm{Tr}(\vert R_S^\prime \vert )=\mathrm{Tr}( P_S( \Gamma_S - \Delta_S )),
\end{aligned}
\end{equation}
where $P_S$ is some projector operator \cite{1}. When  $n \neq m d_E$, the supports of  $\Gamma_S$ and  $\Delta_S$ overlap. Now, there are two possible cases: First, when  $\mathrm{Tr}( P_S \Delta_S )=0$. So, the projector $P_S$ does not span the whole support of $\Gamma_S$. Therefore
\begin{equation}
\label{eq:30}
\begin{aligned}
 D(\rho_S^\prime , \sigma_S^\prime)=\mathrm{Tr}( P_S \Gamma_S ) < \mathrm{Tr}(  \Gamma_S ) = \sum_{i=1}^n \mu_i^{(+)}.
\end{aligned}
\end{equation}
Second, when  $\mathrm{Tr}( P_S \Delta_S )>0$. So
\begin{equation}
\label{eq:31}
\begin{aligned}
 D(\rho_S^\prime , \sigma_S^\prime)=\mathrm{Tr}( P_S( \Gamma_S - \Delta_S )) \qquad\qquad\qquad \quad  \    \\ 
 < \mathrm{Tr}( P_S \Gamma_S ) \leq \mathrm{Tr}(  \Gamma_S ) = \sum_{i=1}^n \mu_i^{(+)}.
\end{aligned}
\end{equation}
Consequently, when $n \neq m d_E$, we never achieve the upper bound given in Eq. \eqref{eq:24}.

In summary, we have proved the following Proposition:
\begin{propo}
\label{pro2}
One can find a unitary evolution $U$, for the whole system-environment, such that the equality sign holds in inequality \eqref{eq:9}, if and only if the number of positive eigenvalues of $R=\rho_{S}\otimes \rho_{E}- \rho_{SE}$  is $n = m d_E$, for some integer $0<m<d_S$.
 \end{propo}

If some eigenvalues of $R$ are zero, we can add them appropriately to the sets    $\lbrace\mu_i^{(+)}\rbrace$ or $\lbrace\mu_i^{(-)}\rbrace$ to achieve the condition $n = m d_E$.
Note that, since $R$ is the difference of two density operators, its eigenvalues falls into the interval $[-1, 1]$. In addition, as $N=d_S d_E$ increases, we expect that a greater number of eigenvalues of a density operator become almost zero, since the sum of the eigenvalues must add up to $1$. Thus, as $N$ increases, we expect that more eigenvalues of $R$ become almost zero. Therefore, even if we cannot achieve the condition $n = m d_E$, we can add those members of $\lbrace\mu_i^{(-)}\rbrace$ which are almost zero to  $\lbrace\mu_i^{(+)}\rbrace$ such that we approach the upper bound in Eq. \eqref{eq:24}.
 
As the final remark, note that we can readily improve the applicability of Proposition \ref{pro2}. 
Consider two arbitrary initial states of the system-environment $\rho_{SE}$ and  $\sigma_{SE}$. So,  using the fact that the trace distance is contractive under the partial trace \cite{1},
 we have 
\begin{equation}
\label{eq:32}
\begin{aligned}
 D(\rho_S^\prime , \sigma_S^\prime) \leq  D(\rho_{SE}^\prime , \sigma_{SE}^\prime)=  D(\rho_{SE} , \sigma_{SE}), 
\end{aligned}
\end{equation}
where $\rho_{SE}^\prime=  U \rho_{SE} U^\dagger$ and  $\sigma_{SE}^\prime=  U \sigma_{SE} U^\dagger$ are the final states of the system-environment, after the unitary evolution $U$. In addition, $\rho_S^\prime =\mathrm{Tr}_E(\rho_{SE}^\prime)$ and $\sigma_S^\prime =\mathrm{Tr}_E(\sigma_{SE}^\prime)$ are the corresponding final states of the system.

Now, instead of inequality \eqref{eq:9}, we can consider the general inequality \eqref{eq:32}. Then, we can follow a similar line of reasoning as given in this section to show that inequality \eqref{eq:32} can be saturated if and only if the number of positive eigenvalues of $R=\sigma_{SE} - \rho_{SE}$  is $n = m d_E$.

Therefore, when $n = m d_E$, using an appropriate $U$, we have  $D(\rho_S^\prime , \sigma_S^\prime) =  D(\rho_{SE} , \sigma_{SE})$, which  can be interpreted as the following: The entire initial distinguishability (information) in the system-environment has been flowed into the system.

\section{When the evolution is given by a time-independent Hamiltonian}  \label{sec: F}


Until now our discussion was restricted to the discrete time evolution case, where the unitary  operator $U$  maps the initial state to the final one, as given in Eq. \eqref{eq:1}. The simplest case for which we can achieve results for all the times is when the evolution is governed by a 
time-independent Hamiltonian.

So in this section  we consider the case that the unitary time evolution of the whole  system-environment is  $U=\exp (-iHt)$, where $H$ is a time-independent Hamiltonian (a Hermitian operator on $\cH_S\otimes\cH_E$),   $i=\sqrt{-1}$ and $t$ is the time. 

Consider the Hermitian operator $R$ in Eq. \eqref{eq:13}. Using the Baker-Hausdorff formula, we can expand  $R^{\prime}=U R U^{\dagger}$ as 
\begin{equation}
\label{eq:33}
\begin{aligned}
R^\prime= e^{(-T)} Re^T= R+ [R,T]+\frac{1}{2!}[[R,T],T]  \\ +  \frac{1}{3!}[[[R,T],T],T]+ \dots ,
\end{aligned}
\end{equation}
where $T=iHt$.
 So we have
\begin{equation}
\label{eq:34}
\begin{aligned}
R^\prime_S=\mathrm{Tr}_{E}(R^\prime)=
\mathrm{Tr}_{E}(R)+ \mathrm{Tr}_{E}([R,T]) \qquad \\
+\frac{1}{2!}\mathrm{Tr}_{E}([[R,T],T]) + \dots,
\end{aligned}
\end{equation}
which is a polynomial function of $t$. Since $\mathrm{Tr}_{E}(R)=0$, the operator $R^\prime_S$ is zero at $t=0$. Therefore, if for some $t>0$ we have $R^\prime_S \neq 0$, we conclude that $R^\prime_S$ is nonzero for almost all the times $t$. This is due to the fact that when a polynomial  is not a fixed function, it can be zero only in a discrete set of  times $t_j$, $j=1, 2, \dots$.
Therefore if inequality \eqref{eq:9} succeeds to detect correlations in the initial $\rho_{SE}$ at some time $t$, i.e., if the corresponding $R^\prime_S$ is nonzero at this $t$, then inequality \eqref{eq:9} succeeds for almost all the other times too.

On the other hand, 
 one may encounter  cases for which the correlation in the initial  $\rho_{SE}$ cannot be detected, by local measurements on the system, at any time $t$.  Such an example is given in Ref. \cite{28}.
There
 we have considered a spin (qubit) chain from spin $1$ to spin $\bar{N}$. 
  We have chosen spins $1$ to $\bar{n}-1$ as our system $S$, and spins $\bar{n}$ to $\bar{N}$ as the environment $E$. In addition,  we have chosen the Hamiltonian as $H=\sum_{j=1}^{\bar{N}-1} Z_j \otimes Z_{j+1}$, where $ Z_j $ is the third Pauli operator
of spin $j$.
  
   Consider the case that the initial state of the spin chain is 
 \begin{equation}
\label{eq:35}
\begin{aligned}
\rho_{SE}=\rho_{S\tilde{E}}\otimes \rho_{\bar{n}},
\end{aligned}
\end{equation}  
 where $\rho_{S\tilde{E}}$ is an arbitrary state on $\cH_S \otimes \cH_{\tilde{E}}$ ($\tilde{E}$ denotes all the spins in the environment $E$, except  spin $\bar{n}$, and $\cH_{\tilde{E}}$ is the corresponding Hilbert space),  and $\rho_{\bar{n}}$  
is an arbitrary state of spin $\bar{n}$, in the $xy$ plane of the Bloch sphere \cite{1}:
 \begin{equation}
\label{eq:36}
\begin{aligned}
\rho_{\bar{n}}=\frac{1}{2}(I_{\bar{n}}+r_x X_{\bar{n}} +r_y Y_{\bar{n}}),
\end{aligned}
\end{equation} 
 where $r_x$ and $r_y$  are real coefficients. In addition, $I_{\bar{n}}$,  $X_{\bar{n}}$ and $Y_{\bar{n}}$ denote the identity operator, the  first and the second Pauli operators on spin $\bar{n}$, respectively.
 
 Therefore the Hermitian traceless  operator $R=\rho_{S}\otimes \rho_{E}- \rho_{SE}$ can be decomposed as
 \begin{equation}
\label{eq:37}
\begin{aligned}
R=R^{(0)}\otimes I_{\bar{n}}+ {R}^{(1)}\otimes X_{\bar{n}} + {R}^{(2)}\otimes Y_{\bar{n}},
\end{aligned}
\end{equation} 
 where ${R}^{(\nu)}$ are Hermitian operators on $\cH_S \otimes \cH_{\tilde{E}}$, such that    $\mathrm{Tr}_{\tilde{E}}({R}^{(0)})=0$, since $\mathrm{Tr}_{E}({R})=0$.

Assuming that the experimenter has access only to the system, and not to the environment, we can construct the state $\sigma_{SE}= \mathcal{F}_S \otimes  \mathrm{id}_E (\rho_{SE})$, performing some quantum operation $ \mathcal{F}_S$ on the system. So, using Eq. \eqref{eq:35}, we have $\sigma_{SE}=\sigma_{S\tilde{E}}\otimes \rho_{\bar{n}}$,  where $\sigma_{S\tilde{E}}$ is a state on $\cH_S \otimes \cH_{\tilde{E}}$.  Therefore, for the Hermitian traceless  operator  $\bar{R}=\sigma_{S}\otimes \sigma_{E}- \sigma_{SE}$, we similarly have
\begin{equation}
\label{eq:38}
\begin{aligned}
\bar{R}=\bar{R}^{(0)}\otimes I_{\bar{n}}+ \bar{R}^{(1)}\otimes X_{\bar{n}} + \bar{R}^{(2)}\otimes Y_{\bar{n}},
\end{aligned}
\end{equation} 
 where $\bar{R}^{(\nu)}$ are Hermitian operators on $\cH_S \otimes \cH_{\tilde{E}}$, such that    $\mathrm{Tr}_{\tilde{E}}(\bar{R}^{(0)})=0$.

Using Lemma 2 of Ref. \cite{28}, we can show that, for the operators $R$ and $\bar{R}$ in Eqs. \eqref{eq:37} and \eqref{eq:38}, respectively, we have $R_S^\prime=\mathrm{Tr}_{E}(URU^\dagger)= 0$  and 
$\bar{R}_S^\prime=\mathrm{Tr}_{E}(U\bar{R}U^\dagger)= 0$, for all the times $t$.
Thus, using inequality \eqref{eq:21}, we conclude that inequality \eqref{eq:7}, or inequality \eqref{eq:9}
 which is a special case of  inequality \eqref{eq:7}, can never detect correlations in the initial state $\rho_{SE}$ in Eq. \eqref{eq:35}.

Note that there is a trivial case for which initial correlation cannot  be detected, for any time $t$: When the time evolution operator is factorized as $U=U_S \otimes U_E$, where $U_S$ and $U_E$ are unitary operators on $\cH_S$ and $\cH_E$, respectively, and so the left hand side of inequality \eqref{eq:7} is always zero.
In our case, because of the term $ Z_{\bar{n}-1} \otimes Z_{\bar{n}}$ in the Hamiltonian, $U=\exp (-iHt)$ is not factorized. Even so, correlations in the initial state $\rho_{SE}$ in Eq. \eqref{eq:35} always remain undetectable, at least using the trace distance method.

\section{Summary} \label{sec: G}


Various methods have been introduced to detect correlations between the system and the environment, in the initial state $\rho_{SE}$, by performing local measurements   only on the system. Maybe the simplest and the best one is that introduced in Ref. \cite{13}, which is based on using inequalities \eqref{eq:7} and \eqref{eq:9}.

In this paper showed that this method always succeeds: In Sec. \ref{sec: C}, using  Proposition \ref{pro1} we showed that one can always find a unitary $U$ such that inequality \eqref{eq:9} can detect correlations in the initial state $\rho_{SE}$.
As another application of Proposition \ref{pro1},  we  discussed  how the presence of initial correlation can affect the quantum process tomography.

In Sec.  \ref{sec: D} we  proved the necessary (and sufficient) conditions for applicability of inequalities \eqref{eq:7} and   \eqref{eq:10} to detect correlations. In particular, we showed that one can always find a unitary $U$ such that inequality \eqref{eq:10} can detect correlation in the environment.

Next, in Proposition \ref{pro2} the necessary  and sufficient condition to saturate inequality \eqref{eq:9} was given. In fact, Proposition \ref{pro2} can be applied to the general case, i.e., inequality \eqref{eq:32}, and determines when we can achieve  the whole initial distinguishability.

Finally, in Sec.  \ref{sec: F} we considered the case that the system and the environment interact through a time-independent  Hamiltonian. 
We saw that for this case,  if inequality \eqref{eq:9} succeeds to detect initial correlations  at some time $t$,   it can succeed for almost all the other times too.
On the other hand, we  discussed an example for which choosing the  Hamiltonian $H$ and the initial state $\rho_{SE}$ appropriately and using inequality \eqref{eq:21} results in inequalities \eqref{eq:7} and \eqref{eq:9}  never detecting correlations, even though $U=\exp (-iHt)$ is not factorized.



%
%
%
%


\begin{thebibliography}{1}


  \bibitem{1} M. A. Nielsen and I. L. Chuang, \textit{Quantum Computation and Quantum Information} (Cambridge University Press, Cambridge, England,  2000). 
  
  
 \bibitem{2}  
H.-P. Breuer and F. Petruccione, \textit{The Theory of Open Quantum
Systems} (Oxford University Press, Oxford, 2002). 



\bibitem{3} H.-P. Breuer, E.-M. Laine, J. Piilo, and B. Vacchini, Colloquium: Non-Markovian dynamics in open quantum systems, \href{https://journals.aps.org/rmp/abstract/10.1103/RevModPhys.88.021002} {Rev. Mod. Phys. \textbf{88}, 021002  (2016)}. 
 
 
  \bibitem{4}    H.-P. Breuer, E.-M. Laine, and J. Piilo, Measure for the degree of non-Markovian behavior of quantum processes in open systems, \href{https://journals.aps.org/prl/abstract/10.1103/PhysRevLett.103.210401}  {Phys. Rev. Lett. \textbf{103}, 210401 (2009)}. 
 
  
 \bibitem{5} Á. Rivas, S. F. Huelga, and M. B. Plenio, Entanglement and non-Markovianity of quantum evolutions, \href{https://journals.aps.org/prl/abstract/10.1103/PhysRevLett.105.050403}  {Phys. Rev. Lett. \textbf{105}, 050403 (2010)}.


\bibitem{6} M. J. W. Hall, J. D. Cresser, L. Li, and E. Andersson, Canonical form of master equations and characterization of non-Markovianity, \href{https://journals.aps.org/pra/abstract/10.1103/PhysRevA.89.042120} {Phys. Rev. A \textbf{89}, 042120  (2014)}.



  \bibitem{7} P. Stelmachovic and V. Buzek , Dynamics of open quantum systems initially entangled with environment: Beyond the Kraus representation, \href{http://dx.doi.org/10.1103/PhysRevA.64.062106} {Phys. Rev. A {\bf 64}, 062106 (2001)};  \href{http://dx.doi.org/10.1103/PhysRevA.67.029902} {\textit{ibid}.  {\bf 67}, 029902(E)(2003)}. 




    \bibitem{8} J. M. Dominy, A. Shabani and D. A. Lidar,  A general framework for complete positivity,   \href{http://link.springer.com/article/10.1007/s11128-015-1148-0}
    { Quant. Inf. Process. {\bf 15}, 465 (2016)}.


  \bibitem{9} J. M. Dominy and D. A. Lidar, Beyond complete positivity,  \href{http://link.springer.com/article/10.1007/s11128-015-1228-1}{ Quant. Inf. Process. {\bf 15}, 1349 (2016)}. 


   \bibitem{10} F. Buscemi, Complete positivity, Markovianity, and the quantum data-processing inequality,
 in the presence of initial system-environment correlations, \href{http://dx.doi.org/10.1103/PhysRevLett.113.140502} {Phys. Rev. Lett. {\bf 113}, 140502 (2014)}. 


     \bibitem{11} I. Sargolzahi, Reference state for arbitrary U-consistent
subspace, \href{https://iopscience.iop.org/article/10.1088/1751-8121/aacaaa/meta} {J. Phys. A: Math. Theor. \textbf{51}, 315301 (2018)}.

 \bibitem{12} I. Sargolzahi, Necessary and sufficient condition for the reduced dynamics of an open quantum system interacting with an environment to be linear, \href{https://journals.aps.org/pra/abstract/10.1103/PhysRevA.102.022208} {Phys. Rev. A \textbf{102}, 022208 (2020)}. 


  \bibitem{13}   E.-M. Laine, J. Piilo, and H.-P. Breuer, Witness for initial system-environment correlations in open-system dynamics,   \href{https://iopscience.iop.org/article/10.1209/0295-5075/92/60010} {Europhys. Lett.    \textbf{92}, 60010   (2010)}.


  \bibitem{14}  G. Amato, H.-P. Breuer, and B. Vacchini, Generalized trace distance approach to quantum non-Markovianity and detection of initial correlations,  \href{https://journals.aps.org/pra/abstract/10.1103/PhysRevA.98.012120} {Phys. Rev. A \textbf{98}, 012120 (2018)}. 

 
 
   \bibitem{15}  C.-F. Li, J.-S. Tang, Y.-L. Li, and G.-C. Guo, Experimentally witnessing the initial correlation between an open quantum system and its environment, \href{https://journals.aps.org/pra/abstract/10.1103/PhysRevA.83.064102} {Phys. Rev. A \textbf{83}, 064102 (2011)}.  
  
 \bibitem{16} A. Smirne, D. Brivio, S. Cialdi, B. Vacchini, and M. G. A. Paris, Experimental investigation of initial system-environment correlations via trace-distance evolution, \href{https://journals.aps.org/pra/abstract/10.1103/PhysRevA.84.032112} {Phys. Rev. A \textbf{84}, 032112 (2011)}.  
  
  
  
  \bibitem{17}   M. Gessner and H.-P. Breuer, Detecting nonclassical system-environment correlations by local operations, \href{https://journals.aps.org/prl/abstract/10.1103/PhysRevLett.107.180402} {Phys. Rev. Lett. \textbf{107}, 180402 (2011)}.
  
  
\bibitem{18} H. Ollivier and W. H. Zurek, Quantum discord: A measure of the quantumness of correlations,  \href{https://journals.aps.org/prl/abstract/10.1103/PhysRevLett.88.017901} {Phys. Rev. Lett. \textbf{88}, 017901  (2001)}.     
  
  
  
 
 
  \bibitem{19} D. M. Tong, L. C. Kwek, C. H. Oh, J.-L. Chen, and L. Ma, Operator-sum representation of time-dependent density operators and its applications,   \href{http://dx.doi.org/10.1103/PhysRevA.69.054102}{Phys. Rev. A {\bf 69}, 054102 (2004)}.   
 
  
    \bibitem{23} F. T. Tabesh, S. Salimi, and A. S. Khorashad, Witness for initial correlations among environments,  \href{https://journals.aps.org/pra/abstract/10.1103/PhysRevA.95.052323} {Phys. Rev. A \textbf{95}, 052323  (2017)}.
 
 
 
 \bibitem{20} S. Hagen and M. Byrd, Detecting initial system-environment correlations in open systems,
  \href{https://journals.aps.org/pra/abstract/10.1103/PhysRevA.104.042406} {Phys. Rev. A \textbf{104}, 042406  (2021)}.
  
 \bibitem{21}  E. Chitambar, A. Abu-Nada, R. Ceballos, and M. Byrd,  Restrictions on initial system-environment correlations based on the dynamics of an open quantum system,   \href{https://journals.aps.org/pra/abstract/10.1103/PhysRevA.92.052110}  {Phys. Rev. A \textbf{92,} 052110  (2015)}.
  
  
\bibitem{22}  G. Zhu , D. Qu, L. Xiao, and P. Xue, Experimental detection of initial system–environment
entanglement in open systems,  \href{https://www.mdpi.com/2304-6732/9/11/883} { Photonics  \textbf{9}, 883 (2022)}.
  
  
  

  

  
  
  \bibitem{24}  A. P. Babu, S. Alipour, A. T. Rezakhani, and T. Ala-Nissila, Unfolding correlation from open-quantum-system master equations,  \href{https://arxiv.org/abs/2104.04248} 
  {arXiv:2104.04248 (2021)}.
  
  
\bibitem{26} M. Ringbauer,  C. J. Wood,
 K. Modi, A. Gilchrist,
 A. G. White,
 and A. Fedrizzi, 
 Characterizing quantum dynamics with initial system-environment correlations, 
\href{https://journals.aps.org/prl/abstract/10.1103/PhysRevLett.114.090402} {Phys. Rev. Lett. \textbf{114}, 090402  (2015)}.  
 
   
   
 
   
  
  
  
\bibitem{27} K. Modi, Operational approach to open dynamics and quantifying
initial correlations,  \href{https://www.nature.com/articles/srep00581}
{Sci. Rep. \textbf{2}, 581 (2012)}.  
  
  
 \bibitem{28} I. Sargolzahi, Instantaneous measurement can isolate the information,  \href{https://arxiv.org/abs/2306.09670}  { arXiv:2306.09670 (2023)}.
  


   
 






  

  
  
  
  
  
%
%
%
  
 
%
%
%
%
%
%
%
%
%
%
%
%
%
%
%
%
%
%
%
%
%
%
 
 

 
 


 
 
 
%
 
  
  
  
%
%
%
 
%
%



%
%
%
%
%
%
%
%
%
%
%
%
%
%
%
%
%
%
%
%

%
%

%
%

%
%
%
%
%
%
%
%

%
%
%








  
  
 
 
 

    
    
%
%
%
%
%




%

%
%
%

%
%

%
%
%
%
%
%
%
%


%
%
%










  
%
%
%
%
%
%
%
%
%
%
%
%
%
%
%
%

%
%
%

%
%
%
%
%

%
%
%
%
%
%
%
%
%

%



%
%
%
%
%
%
%
%
%
%
%
%

\end{thebibliography}
\end{document}